\documentstyle[12pt,fleqn]{article}
\catcode`\@=11
\@addtoreset{equation}{section}
\def\theequation{\arabic{section}.\arabic{equation}}
\def\appendix{\renewcommand{\thesection}{\Alph{section}}
\setcounter{section}{0}
              \renewcommand{\theequation}
            {\mbox{\Alph{section}.\arabic{equation}}}\setcounter{equation}{0}}
\def\maketitle{\thispagestyle{empty}\setcounter{page}0\newpage
                \renewcommand{\thefootnote}{\arabic{footnote}}
                  \setcounter{footnote}0}
\renewcommand{\thanks}[1]{\renewcommand{\thefootnote}{\fnsymbol{footnote}}
               \footnote{#1}\renewcommand{\thefootnote}{\arabic{footnote}}}
\newcommand{\preprint}[1]{\hfill{\sl preprint - #1}\par\bigskip\par\rm}
\renewcommand{\title}[1]{\begin{center}\Large\bf #1\end{center}\rm\par\bigskip}
\renewcommand{\author}[1]{\begin{center}\Large #1\end{center}}
\newcommand{\address}[1]{\begin{center}\large #1\end{center}}
\def\dip{\smallskip Department of Mathematics, University of Massachusetts,\\
           Amherst, Massachusetts 01003}
\def\Idip{\address{\dip}}
\newcommand{\email}[1]{e-mail: \sl #1@math.umass.edu\rm}
\newcommand{\femail}[1]{\thanks{\email{#1}}}

\def\babs{\hrule\par\begin{description}\item{Abstract: }\it}
\def\eabs{\par\end{description}\hrule\par\medskip\rm}
\renewcommand{\date}[1]{\par\bigskip\par\sl\hfill #1\par\medskip\par\rm}
\newcommand{\ack}[1]{\par\section*{Acknowledgements} #1}
\newcommand{\s}[1]{\section{#1}}
\renewcommand{\ss}[1]{\subsection{#1}}

\renewcommand{\vec}[1]{{\bf #1}}       
\def\beq{\begin{eqnarray}}    
\def\eeq{\end{eqnarray}}      
\newtheorem{proposition}{Proposition}          
\def\R{{\hbox{{\rm I}\kern-.2em\hbox{\rm R}}}}   
\def\H{{\hbox{{\rm I}\kern-.2em\hbox{\rm H}}}}   
\def\N{{\hbox{{\rm I}\kern-.2em\hbox{\rm N}}}}   
\def\C{{\ \hbox{{\rm I}\kern-.6em\hbox{\bf C}}}} 
\def\Z{{\hbox{{\rm Z}\kern-.4em\hbox{\rm Z}}}}   
\renewcommand{\Re}{\mathop{\rm Re}\nolimits}       
\def\al{\alpha}
\def\be{\beta}
\def\ga{\gamma}

\def\Ga{\Gamma}

\def\citen#1{%
\edef\@tempa{\@ignspaftercomma,#1, \@end, }
\edef\@tempa{\expandafter\@ignendcommas\@tempa\@end}%
\if@filesw \immediate \write \@auxout {\string \citation {\@tempa}}\fi
\@tempcntb\m@ne \let\@h@ld\relax \let\@citea\@empty
\@for \@citeb:=\@tempa\do {\@cmpresscites}%
\@h@ld}
\def\@ignspaftercomma#1, {\ifx\@end#1\@empty\else
   #1,\expandafter\@ignspaftercomma\fi}
\def\@ignendcommas,#1,\@end{#1}
\def\@cmpresscites{%
 \expandafter\let \expandafter\@B@citeB \csname b@\@citeb \endcsname
 \ifx\@B@citeB\relax 
    \@h@ld\@citea\@tempcntb\m@ne{\bf ?}%
    \@warning {Citation `\@citeb ' on page \thepage \space undefined}%
 \else
    \@tempcnta\@tempcntb \advance\@tempcnta\@ne
    \setbox\z@\hbox\bgroup 
    \ifnum\z@<0\@B@citeB \relax
       \egroup \@tempcntb\@B@citeB \relax
       \else \egroup \@tempcntb\m@ne \fi
    \ifnum\@tempcnta=\@tempcntb 
       \ifx\@h@ld\relax 
          \edef \@h@ld{\@citea\@B@citeB}%
       \else 
          \edef\@h@ld{\hbox{--}\penalty\@highpenalty \@B@citeB}%
       \fi
    \else   
       \@h@ld \@citea \@B@citeB \let\@h@ld\relax
 \fi\fi%
 \let\@citea\@citepunct
}
\def\@citepunct{,\penalty\@highpenalty\hskip.13em plus.1em minus.1em}%
\def\@citex[#1]#2{\@cite{\citen{#2}}{#1}}%
\def\@cite#1#2{\leavevmode\unskip
  \ifnum\lastpenalty=\z@ \penalty\@highpenalty \fi 
  \ [{\multiply\@highpenalty 3 #1
      \if@tempswa,\penalty\@highpenalty\ #2\fi 
    }]\spacefactor\@m}
\begin{document}

\preprint{}

\title{The Conformal Anomaly in General Rank 1 Symmetric Spaces and
Associated Operator Product}

\author{A.A. Bytsenko\thanks{email: abyts@fisica.uel.br\,\,\,\,\,
On leave from Sankt-Petersburg State Technical University, Russia},
A.E. Gon\c calves\thanks{email: goncalve@fisica.uel.br}}
\address{Departamento de Fisica, Universidade Estadual de Londrina,
Caixa Postal 6001, Londrina-Parana, Brazil}
\author{and}
\author{F.L. Williams\femail{williams}}
\Idip

\date{September 1997}

\babs

We compute the one-loop effective action and the conformal anomaly associated
with the product $\bigotimes_p{\cal L}_p$ of the Laplace type operators
${\cal L}_p,\,p=1,2$, acting in irreducible rank 1 symmetric spaces of 
non-compact type. The explicit form of the zeta functions and the conformal 
anomaly of the stress-energy momentum tensor is derived.

\eabs
\noindent PACS numbers:\hspace{1cm} 04.62.+v, 04.60.-m, 02.40.Vh\\
Running title:\hspace{1.5cm}The conformal anomaly in rank 1 symmetric spaces
\s{Introduction}

Recently the important role of the multiplicative anomaly has been recognized
in physics \cite{conn88-117-673,conn90-18-29,conn94,kast95-166-633,
kala95-16-327,mick94-93,acke95-06,acke95-52,conn96-53,cham96-01,cham96-56,
mart96-01,eliz96-56,eliz97-60}. The anomaly associated with multiplicative
properties of regularized determinants of (pseudo-) differential operators can
be expressed by means of the non-commutative residue, the Wodzicki residue
\cite{wodz87} (see also Refs. \cite{kont94,kont94-40}). The Wodzicki residue,
 which is the unique extension of the Dixmier trace to the wider class of
(pseudo-) differential operators \cite{conn88-117-673,kast95-166-633}, has been
considered within the non-commutative geometrical approach to the standard
model of the electroweak interactions \cite{conn90-18-29,conn94,conn96-53,
cham96-01,cham96-56,mart96-01} and the Yang-Mills action functional. Some 
recent papers along these lines can be found in Refs. 
\cite{kala95-16-327,acke95-06,
acke95-52}. Wodzicki's formulae have been used also for dealing with the
singularity structure of the zeta functions \cite{eliz96-56} and the commutator
anomalies of current algebras \cite{mick94-93}. 
 
Therefore it is natural to investigate the multiplicative properties of
differential operators as well as properties of their determinants. The 
product 
of two (or more) differential operators of Laplace type can arise in higher
derivative field theories (for example, in higher derivative quantum gravity
\cite{birr82,buch92}). The partition function corresponding to the product of
two elliptic second order differential operators for the simplest $O(2)$
invariant model of self-interacting charged fields in ${\bf R}^4$ 
\cite{bens91-44-2480} has been derived recently in Ref. \cite{eliz97-60}.
The global additive and multiplicative properties of Laplace type operators
acting in irreducible rank 1 symmetric spaces and the explicit form of the
multiplicative anomaly have been derived in Ref. \cite{byts97}. 

Under such circumstances we should note that the conformal deformations of a 
metric and the corresponding conformal anomaly can also play an important role
in quantum theories with higher derivatives. It is well known that evaluation
of the conformal anomaly is actually possible only for even dimensional spaces
and up to now its computation is extremely involved. The general structure of 
such anomaly in curved $d$-dimensional spaces (d even) has been studied in 
Ref. \cite{dese93-309-279}. We briefly mention here analysis related to this 
phenomenon for constant curvature spaces. The conformal anomaly calculation 
for $d$-dimensional sphere can be found in Ref. \cite{cope86-3-431,derg88}. 
The explicit computation of the anomaly (of the stress-energy tensor) for 
scalar and spinor quantum fields in $d$-dimensional compact hyperbolic spaces 
has been carried out in Ref. \cite{byts95-36-5084} (see also Refs. 
\cite{eliz94,byts96-266-1}) using zeta-function regularization and the 
Selberg trace formula techniques.

The purpose of this paper is to analyze a contribution to the effective
action (in general form) and the conformal anomaly associated with the product
$(\bigotimes_p{\cal L}_p)$, where ${\cal L}_p,\, p=1,2$, are the Laplace type
operators acting in general rank 1 symmentric spaces. 
 
The contents of the paper are the following. In Sect. 2 we review the relevant
information on the irreducible rank 1 symmetric spaces of non-compact type
and the spectral zeta function $\zeta(s|{\cal L})$. The explicit form of the
zeta function $\zeta(s|\bigotimes_p{\cal L}_p)$ and its derivative (at $s=0$),
the one-loop effective action $W^{(1)}$, and the conformal anomaly of the 
stress-energy tensor $<T_\mu^{\mu}(x)>$ related to the operator product are 
given in Sect. 3. Finally we end with some conclusions in Sect. 4.

\s{Irreducible Rank 1 Symmetric Space Forms $\Ga\backslash G/K$ and the 
Spectral Zeta Function}

We shall be working with irreducible rank 1 symmetric spaces $M\equiv X=
G/K$ of non-compact type. Thus $G$ will be a connected non-compact
simple split rank 1 Lie group with finite center and $K\subset G$ a
maximal compact subgroup. Let $\Gamma\subset G$ be a discrete, co-compact,
torsion free subgroup.

Let $\chi$ be a finite-dimensional unitary representation of $\Gamma$,
let $\{\lambda_l\}_{l=0}^{\infty}$ be the set of eigenvalues of the
second-order operator of Laplace type $L=-\Delta_{\Gamma}$ acting on
smooth sections of the vector bundle over $\Gamma\backslash X$ induced
by $\chi$, and let $n_l(\chi)$ denote the multiplicity of $\lambda_l$.

We shall need further a suitable regularization of the determinant of an
elliptic differential operator, and shall make the choice
of zeta-function regularization (see Eq. (3.1)). The zeta function associated 
with the operator ${\cal L}\equiv L+b$ has the form
$$
\zeta(s|{\cal L})=\sum_ln_l(\chi)\{\lambda_l+b\}^{-s}\mbox{;}
\eqno{(2.1)}
$$
here $b$ is an arbitrary constant (an endomorphism of the vector bundle over
$\Ga\backslash X$); $\zeta(s|{\cal L})$ is a well-defined analytic function 
for
$\Re s >\mbox{dim}(M)/2$, and can be analytically continued to a meromorphic
function on the complex plane ${\vec C}$, regular at $s=0$. One can define the
heat kernel of the elliptic operator ${\cal L}$ by
$$
\omega_{\Gamma_{p}}(t)\equiv\mbox{Tr}\left(e^{-t{\cal L}_p}\right)=\frac{-1}
{2\pi i}\mbox{Tr}\int_{{\cal C}_0}dze^{-zt}(z-{\cal L}_p)^{-1}\mbox{,}
\eqno{(2.2)}
$$
where ${\cal C}_0$ is an arc in the complex plane ${\vec C}$. By standard
results in operator theory there exist $\epsilon,\delta >0$ such that for
$0<t<\delta$ the heat kernel expansion holds
$$
\omega_{\Gamma}(t;b,\chi)=\sum_{l=0}^{\infty}n_l(\chi)e^{-(\lambda_l+b)t}
=\sum_{0\leq l\leq l_0} a_l
({\cal L})t^{-l}+ O(t^\epsilon)\mbox{.}
\eqno{(2.3)}
$$

The following representations of $X$ up to local isomorphism can be chosen
$$
X=\left[ \begin{array}{ll}
SO_1(n,1)/SO(n)\,\,\,\,\,\,\,\,\,\,\,\,\,\,\,\,\,\,\,\,\,\,\,\,\,\,\,\,\,
\,\,\,\,\,\,\,\,\,(I) \\
SU(n,1)/U(n)\,\,\,\,\,\,\,\,\,\,\,\,\,\,\,\,\,\,\,\,\,\,\,\,\,\,\,\,\,\,\,
\,\,\,\,\,\,\,\,\,\,(II)\\
SP(n,1)/(SP(n)\otimes SP(1))\,\,\,\,\,(III)\\
F_{4(-20)}/Spin(9)\,\,\,\,\,\,\,\,\,\,\,\,\,\,\,\,\,\,\,\,\,\,\,\,\,\,\,\,
\,\,\,\,\,\,\,(IV)
\end{array} \right]
\mbox{,}
\eqno{(2.4)}
$$
where $n\geq 2$, and $F_{4(-20)}$ is the unique real form of $F_4$ (with
Dynkin diagram $\circ-\circ=\circ-\circ$) for which the character
$(\mbox{dim}X - \mbox{dim}K)$ assumes the value $(-20)$ \cite{helg62}.
We assume that if $G_{1}$ or $G_{2}=SO(m,1)$ or $SU(q,1)$ then $m$ is even and
$q$ is odd.

The suitable Harish-Chandra-Plancherel measure is given as follows:
$$
|C(r)|^{-2}=C_{G}\pi rP(r)\tanh \left(a(G)r\right)=
C_{G}\pi\sum_{l=0}^{\frac{d}{2}-1}a_{2l}r^{2l+1}\tanh \left(a(G)r
\right)\mbox{,}
\eqno{(2.5)}
$$
where
$$
a(G)=\left[ \begin{array}{ll}
\pi \hspace{0.5cm}\mbox{for $G=SO_1(2n,1)$}\\
\frac{\pi}{2} \hspace{0.5cm}\mbox{for $G=SU(q,1),\,\,\,\,\,\,\,\, q$ odd}\\
\hspace{0.8cm}\mbox{or $G=SP(m,1),\,\,\,\,\,\, F_{4(-20)}$}
\end{array} \right]
\mbox{,}
\eqno{(2.6)}
$$
while $C_{G}$ is some constant depending on $G$, and where the $P(r)$ are
even polynomials (with suitable coefficients $a_{2l}$) of degree $d-2$ for
$G\neq SO(2n+1,1)$, and of degree $d-1=2n$ for $G=SO_1(2n+1,1)$
\cite{byts96-266-1,will97-38-796}.

\s{Anomaly Related to the Laplace Type Operator Product}

\ss{The One-Loop Effective Action}

In this section we are interested in multiplicative properties 
$\bigotimes {\cal L}_p$ of the second-order operators of Laplace type 
${\cal L}_p, p=1,2$, related with the one-loop effective action in quantum
field theory. We shall assume a $\zeta$-regularization determinants, i.e.
$$
\mbox{det}_\zeta({\cal L}_p)\stackrel{def}{=}\exp\left(-\frac{\partial}
{\partial s}\zeta(s=0|{\cal L}_p)\right)\mbox{.}
\eqno{(3.1)}
$$

To start with, let us recall the general formalism enabling the treatment of 
the one-loop effective action.
Let the data $(G,K,\Ga)$ be as in Sect. 2, therefore $G$ being one of the
four groups of Eq. (2.4). The trace formula holds
\cite{wall76-82-171,will90-242}
$$
\omega_{\Ga}(t;b,\chi)=V
\int_{\bf R}dre^{-(r^2+b+\rho_0^2)t}|C(r)|^{-2}+\theta_{\Ga}(t;b,\chi)\mbox{,}
\eqno{(3.2)}
$$
where, by definition,
$$
V\stackrel{def}{=}\frac{1}{4\pi}\chi(1)\mbox{vol}(\Ga\backslash G)
\mbox{,}
\eqno{(3.3)}
$$
where $\chi$ is a finite-dimensional unitary representation (or a character) of
$\Ga$, and the number $\rho_0$ is associated with the positive restricted
(real) roots of $G$ (with multiplicity) with respect to a nilpotent factor $N$
of $G$ in an Iwasawa decomposition $G=KAN$. One has $\rho_0=(n-1)/2, n, 2n+1,
11$ in the cases $(I)-(IV)$ respectively in Eq. (2.4). Finally the function
$\theta_{\Ga}(t;b,\chi)$ is defined as follows
$$
\theta_{\Ga}(t;b,\chi)\stackrel{def}{=}\frac{1}{\sqrt{4\pi t}}\sum_{\ga\in C_
\Ga-\{1\}}\chi(\ga)t_\ga j(\ga)^{-1}C(\ga)e^{-(bt+\rho_0^2t+t_\ga^2/(4t))}
\mbox{,}
\eqno{(3.4)}
$$
for a function $C(\ga)$,\,\, $\ga\in\Ga$, defined on $\Ga-\{1\}$ by
$$
C(\ga)\stackrel{def}=e^{-\rho_0t_\ga}|\mbox{det}_{n_0}\left(\mbox{Ad}(m_\ga
e^{t_\ga H_0})^{-1}-1\right)|^{-1}\mbox{.}
\eqno{(3.5)}
$$

The notation used in Eqs. (3.4) and (3.5) is the following. Let $a_0, n_0$
denote the Lie algebras of $A, N$. Since the rank of $G$ is 1,
$\dim a_0=1$ by definition, say $a_0={\bf R}H_0$ for a suitable basis vector
$H_0$. One can normalize the choice of $H_0$ by $\beta(H_0)=1$, where
$\beta: a_0\mapsto{\bf R}$ is the positive root which defines $n_0$; for more
detail see Ref. \cite{will97-38-796}. Since $\Ga$ is torsion free, each
$\ga\in\Ga-\{1\}$ can be represented uniquely as some power of a primitive
element $\delta:\ga=\delta^{j(\ga)}$ where $j(\ga)\geq 1$ is an integer and
$\delta$ cannot be written as $\ga_1^j$ for $\ga_1\in \Ga$, \,\,$j>1$ an
integer. Taking $\ga\in\Ga$, $\ga\neq 1$, one can find $t_\ga>0$
and $m_\ga\in K$ satisfying $m_\ga a=am_\ga$ for every $a\in A$ such that $\ga$
is $G$ conjugate to $m_\ga\exp(t_\ga H_0)$, namely for some
$g\in G, \,g\ga g^{-1}=m_\ga\exp(t_\ga H_0)$. For $\mbox{Ad}$ denoting the
adjoint representation of $G$ on its complexified Lie algebra, one can compute
$t_\ga$ as follows \cite{wall76-82-171}
$$
e^{t_\ga}=\mbox{max}\{|c||c= \mbox{an eigenvalue of}\,\, \mbox{Ad}(\ga)\}
\mbox{,}
\eqno{(3.6)}
$$
in case $G=SO_1(m,1)$, with $|c|$ replaced by $|c|^{1/2}$ in the other cases
of Eq. (2.4).

The spectral zeta function associated with the product $\bigotimes {\cal L}_p$
has the form
$$
\zeta(s|\bigotimes_p{\cal L}_p)=\sum_{j\geq 0}n_j\prod_p^2(\lambda_j+b_p)^{-s}
\mbox{.}
\eqno{(3.7)}
$$
We shall always assume that $b_1\neq b_2$, say $b_1>b_2$. If $b_1=b_2$ then
$\zeta(s|\bigotimes{\cal L}_p)=\zeta(2s|{\cal L})$ is a well-known function.
For $b_1,b_2\in{\bf R}$, set $b_{+}\stackrel{def}{=}(b_1+b_2)/2,\,\, b_{-}
\stackrel{def}{=}(b_1-b_2)/2$, thus $b_1=b_{+}+b_{-}$ and $b_2=b_{+}-b_{-}$.

The zeta function can be written as follows \cite{byts97}

$$
\zeta(s|\bigotimes_{p}{\cal L}_p)=(2b_{-})^{\frac{1}{2}-s}\frac{\sqrt{\pi}}
{\Ga(s)}\int_0^{\infty}dt
$$
$$
\times\left[\frac{\chi(1)\mbox{vol}(\Ga\backslash G)}{4\pi}
\int_{\bf R}dre^{-(r^2+b_++\rho_0^2)t}|C(r)|^{-2}+
\theta_\Ga(t)\right]I_{s-\frac{1}{2}}(b_{-}t)t^{s-\frac{1}{2}}\mbox{.}
\eqno{(3.8)}
$$
where $I_{\nu}(z)$ are the Bessel functions.

Then for $\Re s>0$ Fubini's theorem gives
$$
\int_0^{\infty}dt\frac{\chi(1)\mbox{vol}(\Ga\backslash G)}{4\pi}
\int_{\bf R}dre^{-(r^2+b_++\rho_0^2)t}|C(r)|^{-2}I_{s-\frac{1}{2}}(b_{-}t)
t^{s-\frac{1}{2}}
$$
$$
=(2b_{-})^{\frac{1}{2}-s}\frac{\chi(1)\mbox{vol}(\Ga
\backslash G)\Ga(s)}{4\pi^{3/2}}\int_{\bf R}dr|C(r)|^{-2}\prod_p(r^2+B_p)^{-s}
\mbox{,}
\eqno{(3.9)}
$$
for $B_p=\rho^2_0+b_p$.
In order to analyze the last integral in Eq. (3.9) (for the possibility of
a meromorphic continuation) it is useful to rewrite the function $|C(r)|^{-2}$
(see Eq. (2.5)), using the identity $\mbox{tanh}(ar)\equiv 1-2(1+e^{2ar})^
{-1}$. Then one can calculate a suitable integral in terms of the 
hypergeometric
function $F(\al,\be;\ga;z)$, namely
$$
\int_0^{\infty}drr^{2j+1}\prod_{p}(r^2+B_p)^{-s}=
\frac{\sqrt{\pi}\Ga(2s-j-1)j!}{2^{2s}\Ga(s)\Ga(s+\frac{1}{2})}B_1^{-s}
B_2^{j+1-s}\left(\frac{2B_1}{B_1+B_2}\right)^{j+1}
$$
$$
\qquad\qquad\qquad\qquad\quad \times F\left(\frac{j+1}{2},\frac{j+2}{2};
s+\frac{1}{2};\left(\frac{B_1-B_2}{B_1+B_2}\right)^2\right)\mbox{,}
\eqno{(3.10)}
$$
which is a holomorphic function on $\Re s> (j+1)/2$ and admits a meromorphic
continuation to ${\bf C}$ with only simple poles at points $s=(j+1-n)/2,\,\,
n\in{\cal N}$.

For $Re s>\frac{d}{4}$ the explicit meromorphic continuation holds 
\cite{byts97}:
$$
\zeta(s|\bigotimes_{p}{\cal L}_p)=A\sum_{j=0}^{\frac{d}{2}-1}
a_{2j}\left({\cal F}_j(s)-E_j(s)\right)+{\cal I}(s)
\mbox{,}
\eqno{(3.11)}
$$
where
$$
E_j(s)\stackrel{def}{=}4\int_0^{\infty}\frac{drr^{2j+1}}
{1+e^{2a(G)r}}\prod_{p}(r^2+B_p)^{-s}\mbox{,}
\eqno{(3.12)}
$$
which is an entire function of $s$ and
$$
A\stackrel{def}{=}\frac{1}{4}\chi(1) {\rm vol}(\Ga \backslash G)C_G\mbox{,}
\eqno{(3.13)}
$$
$$
{\cal F}_j(s)\stackrel{def}{=}(B_1B_2)^{-s}
\frac{j!\left(\frac{2B_1B_2}{B_1+B_2}\right)^{j+1}F\left(\frac{j+1}{2},
\frac{j+2}{2};s+\frac{1}{2};\left(\frac{B_1-B_2}{B_1+B_2}\right)^2\right)}
{(2s-1)(2s-2)...(2s-(j+1))}\mbox{,}
\eqno{(3.14)}
$$
$$
{\cal I}(s)\stackrel{def}{=}(2b_-)^{\frac{1}{2}-s}\frac{\sqrt{\pi}}{\Ga(s)}
\int_0^{\infty}dt\theta_\Ga(t,b_+)I_{s-\frac{1}{2}}(b_-t)t^{s-\frac{1}{2}}
\mbox{.}
\eqno{(3.15)}
$$

The goal now is to compute the zeta function and its derivative at $s=0$. 
Thus we have
$$
{\cal F}_j(0)=\frac{(-1)^{j+1}}{j+1}\left(\frac{2B_1}{B_1+B_2}\right)^{j+1}
F\left(\frac{j+1}{2},\frac{j+2}{2};
\frac{1}{2};\left(\frac{B_1-B_2}{B_1+B_2}\right)^2\right)
$$
$$
=\frac{(-1)^{j+1}}{2(j+1)}\sum_l^2B_l^{j+1}
\mbox{,}
\eqno{(3.16)}
$$
$$
E_j(0)=4\int_0^{\infty}\frac{drr^{2j+1}}{1+e^{2a(G)r}}=
\frac{(-1)^j}{j+1}(1-2^{-2j-1})\left[\frac{\pi}{a(G)}\right]^{2j+2}
{\cal B}_{2j+2}
\mbox{,}
\eqno{(3.17)}
$$
$$
{\cal I}(0)=0\mbox{,}
\eqno{(3.18)}
$$
where ${\cal B}_{2n}$ are the Bernoulli numbers.

\begin{proposition}

A preliminary form of the zeta function $\zeta(s|\bigotimes_{p}{\cal L}_p)$
at $s=0$ is
$$
\zeta(0|\bigotimes_{p}{\cal L}_p)=A\sum_{j=0}^{\frac{d}{2}-1}
\frac{(-1)^{j+1}}{2(j+1)}a_{2j}\left\{\sum_l^2B_l^{j+1}+
(2-2^{-2j})\left[\frac{\pi}{a(G)}\right]^{2j+2}
{\cal B}_{2j+2}\right\}
\mbox{.}
\eqno{(3.19)}
$$
\end{proposition}

\begin{proposition}

The derivative of zeta function at $s=0$ has the form:
$$
\zeta'(0|\bigotimes_{p}{\cal L}_p) = A\sum_{j=0}^{\frac{d}{2}-1}a_{2j}\sum_l^4
{\cal E}_l\mbox{,}
\eqno{(3.20)}
$$
where
$$
{\cal E}_1=j!(B_1^{j+1}+B_2^{j+1})\sum_{k=0}^j\frac{(-1)^{k+1}}{k!(j-k)!
(j+1-k)!}\mbox{,}
\eqno{(3.21)}
$$
$$
{\cal E}_2=B_2^{j+1}\left(\frac{B_1-B_2}{2B_1}\right)\frac{(-1)^j}{(j+1)!}
\sum_{k=1}^{\infty}\frac{(j+k+1)!}{(k+1)!}\sigma_k\left(\frac{B_1-B_2}{B_1}
\right)^k\mbox{,}
\eqno{(3.22)}
$$
$$
{\cal E}_3=\log (B_1B_2)\frac{(-1)^j}{2(j+1)}(B_1^{j+1}+B_2^{j+1})
-4\int_0^{\infty}\frac{drr^{2j+1}\mbox{log}\left(\frac{r^2+B_1}{r^2+B_2}
\right)}{1+e^{2a(G)r}}\mbox{,}
\eqno{(3.23)}
$$
$$
{\cal E}_4\equiv{\cal I}'(s=0)=T_\Ga(0,b_1,\chi_1)+T_\Ga(0,b_2,\chi_2)\mbox{,}
\eqno{(3.24)}
$$
and
$$
T_\Ga(0,b_p,\chi_p)\stackrel{def}{=}\int_0^{\infty}dt\theta_\Ga(t,b_p)t^{-1},
\,\,\,\,\,\,\,\, \sigma_k\stackrel{def}{=}\sum_{k=1}^n\frac{1}{k}
\mbox{.}
\eqno{(3.25)}
$$
\end{proposition}

After a standard functional integration the contribution to the Euclidean
one-loop affective action can  be written as follows
$$
W^{(1)}=\frac{1}{2}\mbox{log}\mbox{det}(\bigotimes_p{\cal L}_p/\mu^2)=
-\frac{1}{2}[\zeta'(0|\bigotimes_p{\cal L}_p)+\log\mu^2\zeta
(0|\bigotimes_p{\cal L}_p)]
\mbox{,}
\eqno{(3.26)}
$$
where $\mu^2$ is a normalization parameter. As a result we have
$$
W^{(1)}=-\frac{1}{2}A\sum_{j=0}^{\frac{d}{2}-1}a_{2j}\left[\sum_l^4
{\cal E}_l+\log\mu^2({\cal F}_j(0)-E_j(0))\right]\mbox{,}
\eqno{(3.27)}
$$
where ${\cal F}_j(0), E_j(0)$ and ${\cal E}_l$ are given by the formulae
(3.16), (3.17) and (3.21) - (3.24) respectively.

\ss{The Explicit Form of Anomaly}

In this section we start with a conformal deformation of a (pseudo-) Riemannian
metric and the conformal anomaly of the energy stress tensor. It is well known
that (pseudo-) Riemannian metrics $g_{\mu\nu}(x)$ and $\tilde{g}_{\mu\nu}(x)$ 
on a manifold $M$ are (pointwise) conformal if 
$\tilde{g}_{\mu\nu}(x)=\exp(2f)g_{\mu\nu}(x),\,\,\,f\in C^{\infty}({\bf R})$.
For constant conformal deformations the variation of the connected vacuum 
functional (effective action) can be expressed in terms of the generalized zeta
function related to an elliptic self-adjoint operator ${\cal O}$
\cite{birr82}
$$
\delta W=-\zeta(0|{\cal O})\log \mu^2=\int_M dx<T_{\mu\nu}(x)>
\delta g^{\mu\nu}(x)\mbox{,}
\eqno{(3.28)}
$$
where $<T_{\mu\nu}(x)>$ means that all connected vacuum graphs of the stress-
energy tensor $T_{\mu\nu}(x)$ are to be included. Therefore the Eq. (3.28) 
leads to
$$
<T_\mu^\mu (x)>=(\mbox{Vol}M)^{-1}\zeta(0|{\cal O})\mbox{.}
\eqno{(3.29)}
$$
In the case of sphere $S^d$ of unit radius we have for example
$(\mbox{Vol}S^d)=2\pi^{(d+1)/2}/\Ga((d+1)/2)$, while the Eqs. (3.2) and (3.3)
give $VC_G\pi=(\mbox{Vol} M)[(4\pi)^{d/2}\Ga(d/2)]^{-1}$ (see for detail Ref. 
\cite{byts96-266-1}). As a result we have $(\mbox{Vol} M)=A(4\pi)^{d/2}
\Ga(d/2)$.

The formulae (3.11), (3.16), (3.17) and (3.18) give an explicit result for 
the conformal anomaly, namely
$$
<T_\mu^\mu(x)>_{({\cal O}=\bigotimes{\cal L}_p)}=\frac{1}{(4\pi)^{d/2}\Ga(d/2)}
\sum_{j=0}^{\frac{d}{2}-1}\frac{(-1)^{j+1}}{2(j+1)}a_{2j}
$$
$$
\times\left\{\sum_l^2B_l^{j+1}
+(2-2^{-2j})\left[\frac{\pi}
{a(G)}\right]^{2j+2}{\cal B}_{2j+2}\right\}
\mbox{,}
\eqno{(3.30)}
$$
where $d$ is even.

For $B_1=B_2=B$ the anomaly (3.30) is associated with Laplace type operator
${\cal L}=L+b$ and has the form
$$
<T_\mu^\mu(x)>=\frac{1}{(4\pi)^{d/2}\Ga(d/2)}\sum_{j=0}^{\frac{d}{2}-1}
\frac{(-1)^{j+1}}
{2(j+1)}a_{2j}\left\{B^{j+1}+(2-2^{-2j})\left[\frac{\pi}{a(G)}\right]^{2j+2}
B_{2j+2}\right\}
\mbox{.}
\eqno{(3.31)}
$$
Note that for minimally coupled scalar field of mass $m$, $B=\rho_0^2+m^2$.

The simplest case is, for example $G=SO_1(2,1)\simeq SL(2,{\bf R})$; besides
$X=H^2$ is a two-dimensional real hyperbolic space. Then we have $\rho_0^2=
\frac{1}{4}, a_{20}=1, C_G=1, a(G)=\pi, |C(r)|^{-2}=\pi r\tanh (\pi r)$, and
finally
$$
<T_\mu^\nu(x\in H^2)>=-\frac{1}{4\pi}(b+\frac{1}{3})
\mbox{.}
\eqno{(3.32)}
$$

For real $d$-dimensional hyperbolic space $C_G=[2^{d-2}\Ga(d/2)]^{-2}$, while
the scalar curvature is $R(x)=-d(d-1)$. For the conformally invariant scalar
field we have $B=\rho_o^2+R(x)(d-2)/[4(d-1)]$. As a consequence, for all
constant curvature spaces $B_1=B_2$; for hyperbolic spaces $B=\frac{1}{4}$ and
$$
<T_\mu^\mu(x\in H^d)>=\frac{1}{(4\pi)^{d/2}\Ga(d/2)}\sum_{j=0}^{\frac{d}{2}-1}
\frac{(-1)^{j+1}}{j+1}a_{2j}\left\{2^{-2j-2}+(1-2^{-2j-1})B_{2j+2}\right\}
\mbox{.}
\eqno{(3.33)}
$$
Thus in conformally invariant scalar theory the anomaly of the stress 
tensor coincides with one associated with  operator product. This 
statement holds not only for hyperbolic spaces considered above but for all 
constant curvature manifolds as well.

\s{Conclusions}

In this paper the one-loop contribution to the effective action (3.27) and the
conformal anomaly of the stress-energy momentum tensor (3.30) related to the
operator product have been evaluated explicitly. In addition we have considered
the product $\bigotimes_p{\cal L}_p$ of Laplace type operators 
${\cal L}_p$ acting in irreducible rank 1 symmetric spaces. As an example the 
conformal anomaly has been computed for real $d$-dimensional hyperbolic spaces.

We have shown that for the class of constant curvature manifolds the conformal
anomalies associated with the Laplace type operator ${\cal L}$ and the product
$\bigotimes_p{\cal L}_p$  coincide. Our formulae can be generalized to the 
case of transverse and traceless tensor fields and spinors in real hyperbolic
spaces (see, for example, \cite{camp92-148-283,camp93-47-3339,camp94-35-4217}).
Indeed in the case of the spin-1 field (vector field theory) for instance we
should draw attention to the fact that the Hodge-de Rham operator $-(d\delta+
\delta d)$ acting on co-exact one-forms corresponds to the mass operator
$(L+d-1)g_{\mu\nu}(x)$. The eigenvalues of this operator are $\lambda_l+
(\rho_0-1)^2$, and for the Proca field of mass $m$ we have $B=(\rho_0-1)^2+
m^2$. Finally we have also computed the anomaly in a simple situation, namely
for the conformally invariant scalar fields. Our result (3.33) coincides 
with the quantum correction reported in Ref. \cite{byts95-36-5084} for compact
hyperbolic spaces. Recently the conformal anomaly of dilaton coupled matter 
in four dimensions has been calculated in Refs. \cite{noji97-09,noji97-43}. 
It would be of great interest to generalize our results to the dilaton 
dependent trace anomaly.

An extension of the above evaluation of the effective action and the conformal
anomaly for higher spin fields seems to us certainly feasible. The analysis of 
multiplicative properties of Laplace type operators and related zeta functions
will be interesting in view of future applications to concrete problems in
quantum field theory and for mathematical applications as well.

\ack{ A.A. Bytsenko wishes to thank CNPq and the Department of Physics of 
Londrina University for financial support and kind hospitality. The research 
of A.A. Bytsenko was supported in part by Russian Foundation for Fundamental 
Research Grant No. 95-02-03568-a and by Russian Universities Grant 
No.~95-0-6.4-1.}

\end{document}